\documentclass[sigconf]{acmart}

\usepackage{bbding}
\usepackage{hyperref}
\usepackage{enumitem}
\usepackage{xcolor}

\AtBeginDocument{%
  \providecommand\BibTeX{{%
    \normalfont B\kern-0.5em{\scshape i\kern-0.25em b}\kern-0.8em\TeX}}}

\setcopyright{acmcopyright}
\copyrightyear{2018}
\acmYear{2018}
\acmDOI{XXXXXXX.XXXXXXX}

\acmConference[ACM Multimedia Systems '23]{Make sure to enter the correct
  conference title from your rights confirmation emai}{June 07--10,
  2023}{Vancouver, Canada}
\acmPrice{15.00}
\acmISBN{978-1-4503-XXXX-X/18/06}

\copyrightyear{2023}
\acmYear{2023}
\setcopyright{acmlicensed}\acmConference[MMSys '23]{Proceedings of the 14th ACM Multimedia Systems Conference}{June 7--10, 2023}{Vancouver, BC, Canada}
\acmBooktitle{Proceedings of the 14th ACM Multimedia Systems Conference (MMSys '23), June 7--10, 2023, Vancouver, BC, Canada}
\acmPrice{15.00}
\acmDOI{10.1145/3587819.3592551}
\acmISBN{979-8-4007-0148-1/23/06}
\begin{document}

\title{FSVVD: A Dataset of Full Scene Volumetric Video}

\author{Kaiyuan Hu$^{1,2 \dagger}$, Yili Jin$^{1,2 \dagger}$, Haowen Yang$^{1,5}$, Junhua Liu$^{1}$, Fangxin Wang$^{2,1,3,4}$}\authornote{Fangxin Wang is the corresponding author.\\$\dagger$Both authors contributed equally to this research.}
\affiliation{%
 \institution{$^1$The Future Network of Intelligence Institute, The Chinese University of Hong Kong, Shenzhen\\$^2$School of Science and Engineering, The Chinese University of Hong Kong, Shenzhen\\$^3$The Guangdong Provincial Key Laboratory of Future Networks of Intelligence\\$^4$Peng Cheng Laboratory,\
 $^5$Versee Inc.
 }
 \streetaddress{}
  \city{}
  \country{}}
\email{{kaiyuanhu,yilijin,haowenyang,junhualiu}@link.cuhk.edu.cn, wangfangxin@cuhk.edu.cn}




\renewcommand{\shortauthors}{Kaiyuan Hu, Yili Jin, Haowen Yang, Junhua Liu, Fangxin Wang}

\begin{abstract}

Recent years have witnessed a rapid development of immersive multimedia which bridges the gap between the real world and virtual space. Volumetric videos, as an emerging representative 3D video paradigm that empowers extended reality, stand out to provide unprecedented immersive and interactive video watching experience. Despite the tremendous potential, the research towards 3D volumetric video is still in its infancy, relying on sufficient and complete datasets for further exploration. However, existing related volumetric video datasets mostly only include a single object, lacking details about the scene and the interaction between them. In this paper, we focus on the current most widely used data format, point cloud, and for the first time release a full-scene volumetric video dataset that includes multiple people and their daily activities interacting with the external environments. Comprehensive dataset description and analysis are conducted, with potential usage of this dataset. The dataset and additional tools can be accessed via the following website:\ \href{}{https://cuhksz-inml.github.io/full\_scene\_volumetric\_video\_dataset/}.

\end{abstract}

\ccsdesc[500]{Information systems~Multimedia databases}
\ccsdesc[500]{Computing methodologies~Volumetric models}

\keywords{Volumetric video, datasets, XR}

\begin{teaserfigure}
\vspace{-15pt}
  \includegraphics[width=\textwidth]{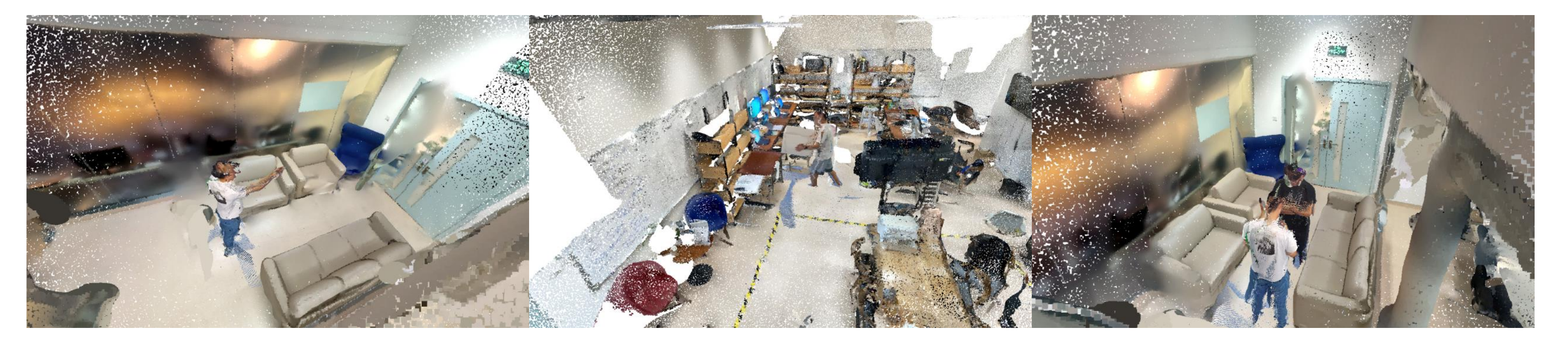}
  \vspace{-20pt}
  \caption{Sample frames of the released dataset sequence}
  \label{title}
\end{teaserfigure}

\maketitle

\section{Introduction}

Recent years have witnessed a rapid development of immersive multimedia which bridges the gap between the real world and virtual space. Volumetric videos, as an emerging representative 3D video paradigm that empowers extended reality, stand out to provide unprecedented fully immersive, and 6 degree-of-freedom (DoF) interactive video watching experience. According to industry research provider "The Insight Partners", the volumetric video market is expected to grow from US\$1,898.7 million in 2021 to US\$9,685.7 million by 2028.

\begin{table*}[t]
\vspace{-10pt}
    \centering
    \caption{\textbf{Comparisons with related datasets
    }}
    \vspace{-10pt}
    \resizebox{\textwidth}{!}{
    \begin{tabular}{|c|c|c|c|c|c|c|c|c|}
    \hline
        \textbf{} & \textbf{Point Clouds} & \textbf{Multi-person} & \textbf{Multi-view Depth} & \textbf{3D-Mesh} & \textbf{Interactive Objects} & \textbf{Full Scenes} & \textbf{\#Videos} & \textbf{\#Categories} \\ \hline
        \textbf{CWIPC-SXR \cite{cwipc}} & \checkmark & \checkmark & \checkmark & - & \checkmark & - & 21 & 4 \\ \hline
        \textbf{HUMAN4D \cite{human4d}} & \checkmark & \checkmark & \checkmark & \checkmark & \checkmark & - & 19 & 3 \\ \hline
        \textbf{HUMBI \cite{HUMBI}} & - & - & - & \checkmark & - & - & 700+ & N/A \\ \hline
        \textbf{Volograms \& V-SENSE \cite{Volograms}} & \checkmark & - & - & - & - & - & 3 & 3 \\ \hline
        \textbf{Ours} & \checkmark & \checkmark & \checkmark & - & \checkmark & \checkmark & 26 & 4 \\ \hline
    \end{tabular}}
    \vspace{-10pt}
\end{table*}
\label{related_datasets}

The key difference between volumetric video compared with traditional 2D flat video~\cite{DBLP:conf/mm/JinL0C22,10090453} lies in the 3D representation, where the commonly used formats are point cloud, mesh, voxel, and the recent implicit neural representation. Among all these representations, point cloud is currently the most popular due to its simplicity and easy deployment~\cite{cav3}. Such video transition from 2D to 3D volumetric video actually brings quite a few new challenges, such as the vast data size, fast dynamic 6 DoF interaction, ultra-low latency requirement, client-side lightweight and smooth playback, etc. Since the research towards 3D volumetric video is still in its infancy, a sufficient and complete dataset is in urgent need in this community for further exploration.


There are a few existing datasets on volumetric video, yet they either only focus on single-person capture and reconstruction or lack the interaction between the target person and the surrounding environment. Thus, we are motivated to collect a comprehensive and complete dataset to facilitate further research.


In this paper, we release a publicly available point-cloud-based Full Scene Volumetric Video Dataset (FSVVD), which contains more than 30 different daily life scenarios. To our best knowledge, FSVVD is the first full-scene volumetric video dataset containing both multiple target people and the interacting environment. We also introduce the detailed capture procedure and release our designed processing tools. A comprehensive analysis of the impact of factors on the volumetric dataset quality is further conducted, including the factors of object movement, scene complexity, and illumination condition. We find that object movement has more impact on volumetric video quality causing ghost images.
We finally point out some potential usage based on our dataset.

The rest of the paper is organized as follows. Section ~\ref{existing_datasets} introduces previous works of related datasets. Section ~\ref{capture_procedure} describes the whole pipeline and principle of capturing volumetric video data. In Section \ref{dataset}, the content of FSVVD is introduced, and in Section \ref{data_analysis}, we make a comprehensive analysis of the impact of factors on the volumetric dataset quality. Finally, we point out some potential usage of our dataset and make conclusions in Section \ref{uses} and Section \ref{conclusions}.


\section{Existing Datasets}\label{existing_datasets}

Over the past few years, both academia and industry have shown great interest in capturing volumetric video data using a variety of RGB-based or depth-based methods. Such volumetric video data promote the research over immersive multimedia technologies like inter-frame interpolation ~\cite{VoluSR}, compression \& decompression, and streaming. We make a brief overview of the related datasets, and summarize the features and modalities by making comparisons, which is illustrated in Table ~\ref{related_datasets}.

 Some previous works only use conventional RGB cameras for capture.
 HUMBI ~\cite{HUMBI} is a publicly available multi-view image dataset, which provides human body expressions including gaze, face, hand, body, and garment. The capture exploited 107 synchronized HD cameras and more than 700 subjects across gender, ethnicity, age, and style. The dataset provides 3D mesh models along with multi-view image streams. And it aims to exert an influence on learning and reconstructing a complete human model and complementing the existing datasets of human body expressions with limited views and subjects. However, depth cameras are not involved in the capture process, thus multi-view depth data is absent.

 Another available volumetric sequence dataset is Volograms \& V-Sense dataset ~\cite{Volograms}. The dataset includes three sequences featuring three different characters, each is captured using 12 HD cameras with a different purpose and application in mind and has different characteristics in terms of texture and movement. Unfortunately, this dataset contains very few movements and scenarios.

Conversely, some works exploit depth cameras in the capture process in order to obtain volumetric data composed of point clouds.
 CWIPC-SXR ~\cite{cwipc} consisting of 21 sequences, provides a dataset for social XR scenarios captured by commodity depth sensors. In such a release, human interaction in social XR scenarios is depicted in the aspects of four use cases: “Education and Training”, “Healthcare”, “Communication and Social interactions”, and “Performance and Sports”, which only focus on the movement of individuals in practical scenarios without the full scene.
 
 Meanwhile, Human4D ~\cite{human4d} dataset contains simultaneously captured human activities using both motion capture devices and depth sensors, such a motion capture method utilized 24 motion capture cameras to achieve higher accuracy. And the method requires actors to wear a pure black uniform with color dots to locate key skeleton points to obtain reference animation of the actors, which is not practical in real-life scenarios. The dataset provides a reproduction of human physical, daily and social activities. Though the captured data are of high accuracy, such capture setups are difficult to deploy in practical use and they lack interaction in the full scene as well.
 
 Regrettably, all of the above datasets lack human interaction with objects in a full scene, and such a feature provides opportunities in tackling challenging tasks of multi-object occlusion \cite{Vivo}and inter-frame interpolation in immersive multimedia. However, FSVVD fills the gap by providing volumetric video data that not only depicts human interaction with objects but also the full scene involved, which aims to energize the research on immersive multimedia topics including volumetric video streaming and real-time telepresence.

\begin{figure*}[t]
\centering
\vspace{-12pt}
\includegraphics[width=1.7\columnwidth,trim=0 10 0 20, clip]{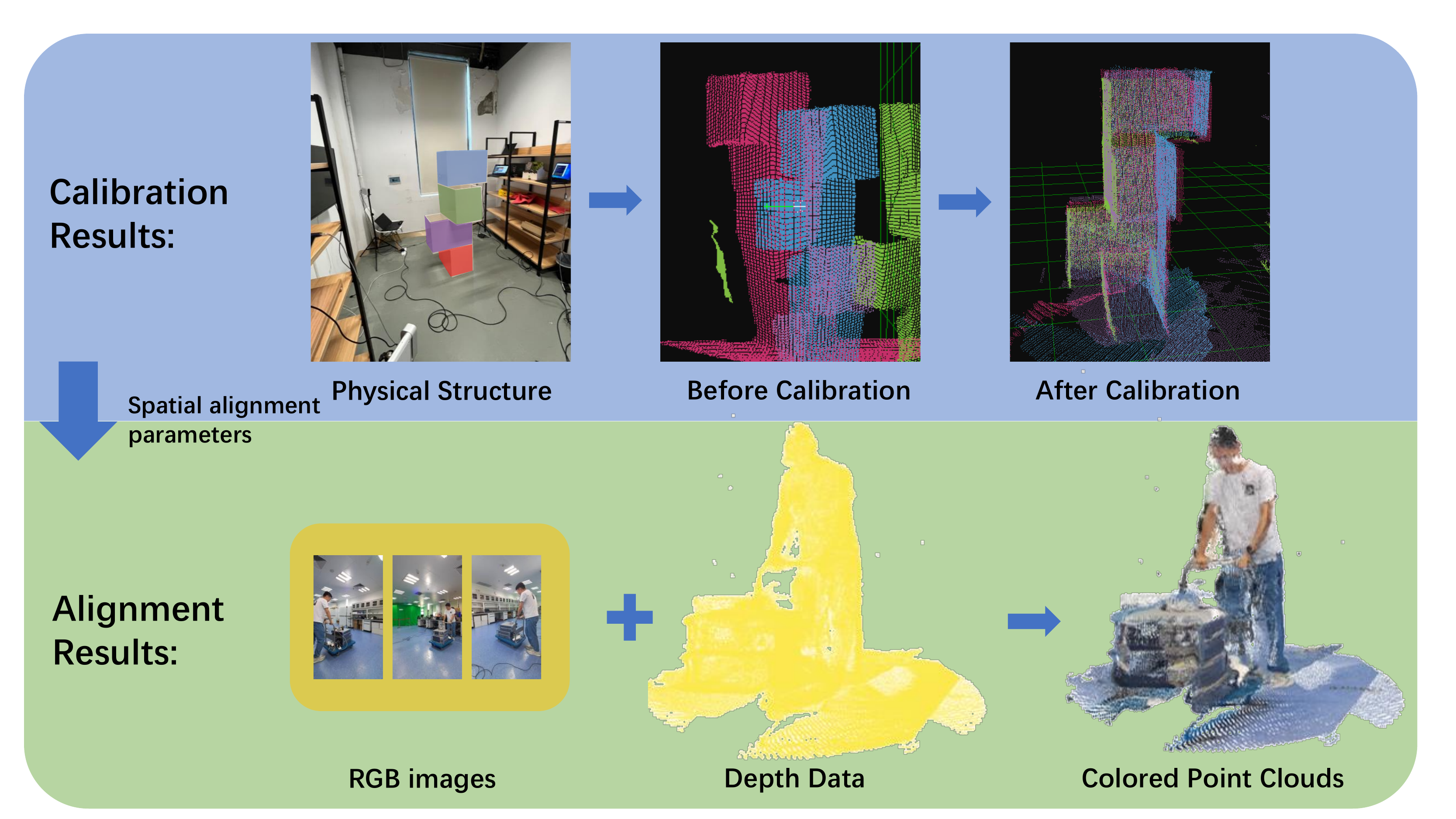}
 \vspace{-10pt}
 \caption{The pipeline of data spatial alignment: calibration and alignment}
 \vspace{-12pt}
 \label{calibration_alignment}
\end{figure*}

\section{capture procedure}\label{capture_procedure}
In the data acquisition process, we design a capture tool that enables multiple devices to capture the data simultaneously. Plus, a data processing tool to convert the raw data in binary into the readable file format of .jpg, .ply, and .pgm. For the capture setup, we build the capture environment with 6 commodity Microsoft Azure Kinect devices in several daily scenes including an office, a warehouse, a meeting room, and a kitchen. Each Azure Kinect device is connected to a client PC for data pre-processing and transmission purposes, and all of the client PCs are connected to a high-performance host PC which is responsible for data processing. The capture environment is illuminated by 2 lamps according to needs, each having a power of 28W.

\subsection{Software Setup}
For the software setup, we build a volumetric capture tool based on the Microsoft Azure Kinect APIs, and a data processing tool to convert the captured raw data in binary into .ply format for further processing, which is fitting the color data to depth data. The data in the interested region can also be extracted by filtering using the data processing tool.

\subsection{Hardware Setup}
For the hardware setup, we only use commodity devices to build the capture environment. We use 6 Microsoft Azure Kinect devices to capture color and depth data, each device is connected to a client PC for data pre-processing and transmission to the host. The client PCs are commodity laptops, which are connected to the sensors using USB 3.0 cables. While the host is a high-performance PC running Windows 10 Pro on an Intel i7-10700 GPU with an NVIDIA GeForce RTX 3060 GPU and 16GB RAM that is responsible for converting raw data in binary into colored point clouds data in the format of .ply. Each of the Azure Kinect devices is uniformly placed facing inwards perimetrically around the center of the capturing region with a radius of 1.75 meters. In order to capture objects at a closer distance by exploiting the sensor's wide horizontal field of view~ \cite{kinectfov}, all of the Azure Kinect devices are placed vertically. In addition, to attain better adjustability, we attach each device on a tripod and set the height and angle of the sensors according to different capture environments.

\subsection{Synchronization}
We exploit hardware synchronization in order to synchronize each sensor in data capturing. The 6 sensors are interconnected in a special topology of daisy-chain by 3.5mm audio cables, with 1 `master' device and 5 `slave' devices. During capturing, the master device will trigger other slave devices successively with an interval of 30 microseconds for each frame, enabling the camera array to take the shot simultaneously. Since the data flow bandwidth of a single device is as much as 80 Mbps, to support real-time data transmission between the host and client PC, we use a network switch of 1000 Mbps bandwidth and multiple Ethernet wires to connect the host and the client PCs.

\subsection{Calibration}
\label{Calibration}
Multi-sensor spatial alignment is essential in exploiting multiple depth sensors. Typically, such a procedure is achieved via marker-based methods, where all views align with respect to some markers and are then registered among themselves. Traditional methods involve moving a checkboard pattern within the capturing region or in the case of retro-reflective markers, a wand, while in our implementation of spatial alignment of Azure Kinects, we use a structure-based method without markers but instead a physical structure assembled by 4 commonly found boxes. During the calibration process, we assemble the alignment structure in the center of the capturing region, such that the structure is fully exposed to all of the sensors. For that, we rely on data-driven correspondence establishment for the initial matching ~\cite{StructureNet,alignment}, and global optimization for estimating a solution with respect to the coordinate system of the structure. The illustration of the calibration process is shown in the blue region of Figure ~\ref{calibration_alignment}.

\subsection{Alignment}
\label{Alignment}
The captured data is originally in binary format, we build a data processing tool in order to convert the data into the displayable format of .ply.
The process of alignment is illustrated in the green region of Figure ~\ref{calibration_alignment}.
By exploiting the spatial alignment parameters from the calibration process, the captured RGB images are aligned with the depth data to obtain the colored point clouds.

\begin{table*}[!ht]
\vspace{-10pt}
    \centering
    \caption{\textbf{Description of the dataset: Each sequence is recorded separately according to different screenplays, and the frame numbers are provided. The information in detail of each screenplay is provided in the screenplay.txt in each sequence package.
    }}
    \vspace{-10pt}
    \begin{tabular}{lllllll}
    \hline
         Category & Screenplay & Actors & Frame Number  & Scene & ~ \\ \hline
        ~ & Presenting & 1 Male & 191 & meeting room & ~ \\ 
         & Answering questions & 1 Male & 339 & warehouse & ~ \\ 
         Education & Writing on a iPad & 1 Male & 330 & warehouse & ~ \\ 
         & Talking in front of whiteboard & 1 Male & 154 & meeting room & ~\\
         & Clean whiteboard & 1 Male & 252 & meeting room & ~\\
         \cline{1-5}
        ~ & Lift box & 1 Male & 256 & office & ~ \\ 
         Exercise & Pulling trolley & 1 Male & 222 & warehouse & ~ \\
        & Rope skipping & 1 Male  & 252 & warehouse & ~ \\ 
         \cline{1-5}
        ~ & Sit down & 1 Male & 220 & office & ~ \\ 
         & Stand up & 1 Male & 132 & office & ~ \\ 
         & Tie shoes & 1 Male & 194 & office & ~ \\ 
        & Drinking & 1 Male & 316 & office & ~ \\ 
         & Sweep the floor & 1 Male & 538 & warehouse & ~ \\ 
         & Chatting between 2 person & 2 Male & 191 & warehouse & ~ \\ 
          Daily life & Use iPad & 1 Male & 135 & warehouse & ~\\
         & Use laptop & 1 Male & 399 & office & ~\\
         & Interview & 2 Male & 207 & office & ~\\
         & Discussing & 2 Male & 640 & meeting room & ~\\
         & Use Quest Pro & 1 Male & 180 & kitchen & ~\\
         & News interview & 1 Male + 1 Female & 2192 & meeting room & ~\\
         & Walk around & 1 Male & 127 & warehouse & ~\\                  
         \cline{1-5}
        & Vive interact  & 1 Male & 383 & warehouse & ~ \\ 
        & Use Hololens & 1 Male & 226 & office & ~ \\ 
        Entertainment & Playing violin & 2 Female & 1863 & empty room & ~ \\ 
        & Play VR action games & 1 Male & 434 & kitchen & ~\\
        & Play guitar & 1 Male & 212 & office & ~\\
        \hline
    \end{tabular}
    \label{list_of_screenplay}
    \vspace{-12pt}
\end{table*}

\section{Dataset}\label{dataset}

In order to reproduce real-life scenarios as realistically as possible, we design 4 categories of volumetric data sequences with respect to 4 real-life scenarios: `Education', `Exercise', `Daily life', and `Entertainment'. Each of the screenplays is thoroughly designed in order to fully depict daily interactive scenarios with detail. The duration of the sequence range from 5 to 88 seconds and the total number of frame range from 127 to 2192. 
To assist further research on volumetric video, we design some challenging props in certain screenplays, for example, small objects like a VR controller are involved in the screenplay `Vive interact' and `Play VR action games'.
The description of the cases is shown in Table ~\ref{list_of_screenplay} in the aspects of screenplay name, actor information, frame number, and scene.

In our recordings, we capture 26 sequences of volumetric videos in total, 21 include single actors and 5 include multiple actors. According to the design of the screenplays, the actors are dressed corresponding to the scene. To preserve the RGB information to the maximum extent, we compensate for the ambient light using two LED photoflood lamps. 

\begin{figure}[t]
\centering
\includegraphics[width=0.8\columnwidth, trim=0 0 0 20, clip]{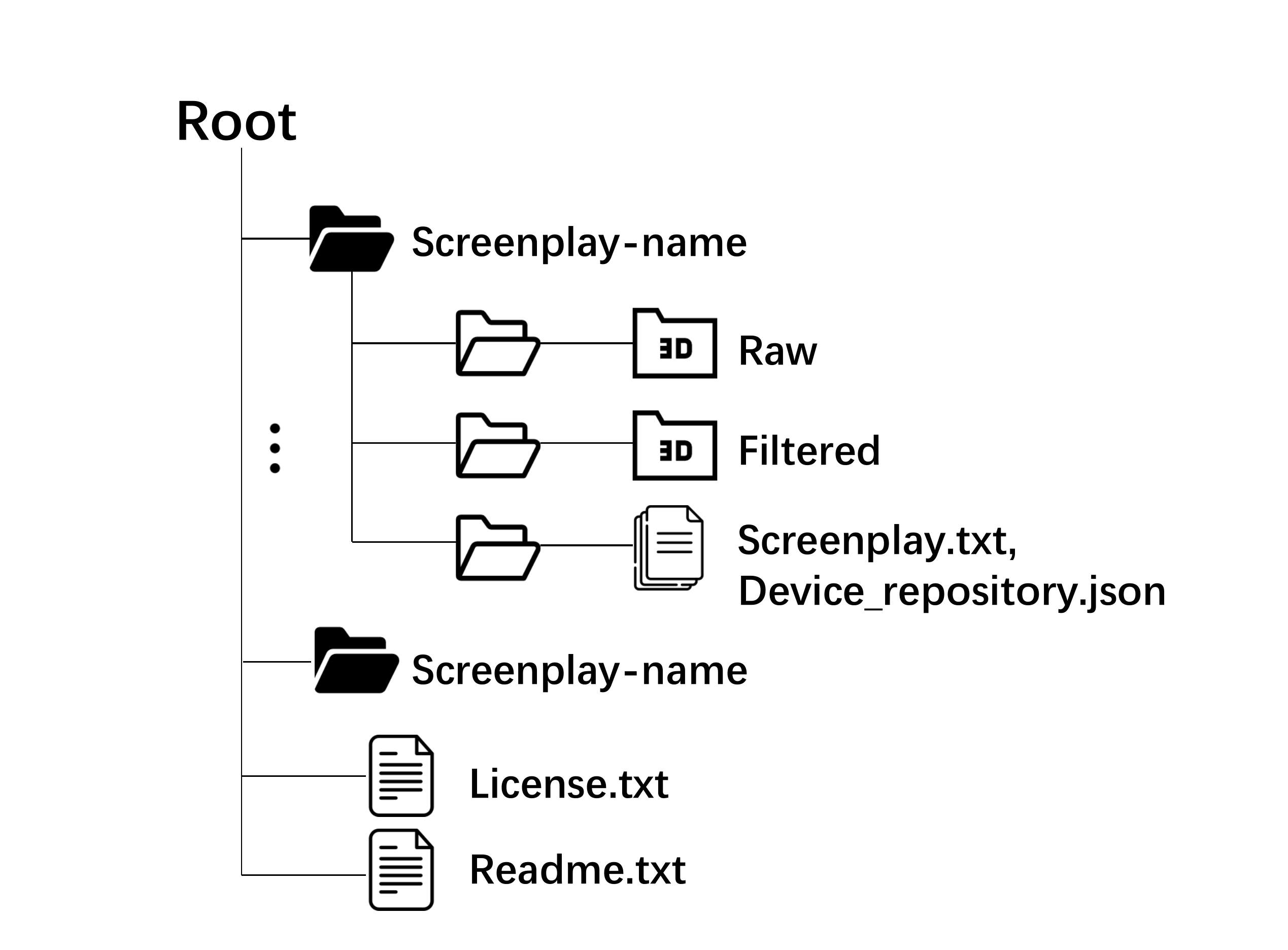}
\vspace{-15pt}
 \caption{Illustration of dataset structure}
 \vspace{-15pt}
 \label{structure}
\end{figure}

The structure of the dataset is described in Figure ~\ref{structure}. For every screenplay folder, there are 3 sub-folders, 2 of which contain the point cloud data, one is the data after filtering, which only focus on the contents in the center of the capture region and the other is the original data that contains the texture of surrounding environments. All of the volumetric sequences are stored in .ply format, the naming convention is: screenplay-name\_frame-number\_Raw/Filtered.ply. (For example, 'Use\_Hololens\_204\_Filtered.ply' represents the 204th frame of the screenplay of `using a Hololens device', with the data filtered by the processing tool.)
The other folder contains supplementary files, including a screenplay.txt file describing the detailed setups of the screenplay, and a device\_repository.json describing the device configurations and the capture setups. 

\section{Data analysis}\label{data_analysis}
Processing and rendering volumetric data is a huge computation drain ~\cite{Lowlatency,Vues}. We conduct several analyses on how certain factors will impact the data quality and computational performance of our proposed dataset in the aspects of capture device setups, object movement, scene complexity, and illumination condition. The affected factors include volumetric video data size, point cloud density, and image detail loss.


\begin{figure}[t]
\centering
\includegraphics[width=\columnwidth]{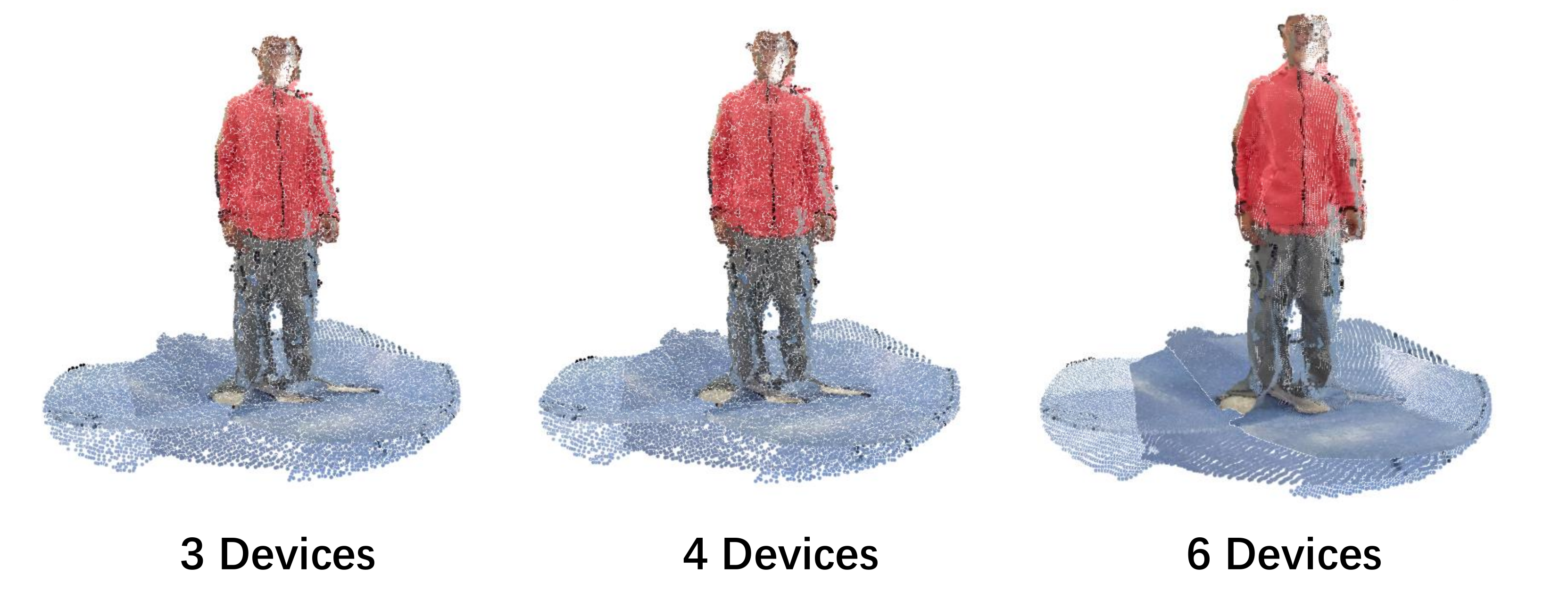}
\vspace{-20pt}
\caption{Different device setups}
\vspace{-20pt}
\label{device}
\end{figure}

\subsection{Device Setups}



Due to the limitation of the sensor's FOV ~\cite{kinectfov}, there will be a loss in detail to a certain extent depending on the number of depth cameras and occlusion conditions, especially in scenarios with multiple objects. In order to assess the impact, we use 3, 4, and 6 sensors to capture the same scenario for the same duration. For the capture process, we keep the capture setups (screenplay, illumination condition, camera configurations) all the same. To get a better observation of the impact, we capture a static scenario with only one actor in this experiment. In each capture, the same actor stays at the same position and makes no movement, and the duration is all set as 50 frames. The experiment result shows: For the same scenario, the average points number per frame is 40210 for 3 devices, 57695 for 4 devices, and 81630 for 6 devices. Based on the results, we summarize the conclusion as follows: The level of detail of the captured data is higher when employing more devices. And a loss of detail will arise in some areas where the components are thin or small (e.g., hand gestures). Accordingly, with the loss of detail, the number of points per frame is lower when fewer cameras are deployed, with more capture devices, this problem could be relieved. To illustrate our findings, several example frames are provided in Figure ~\ref{device}, from which we can see the facial details and surface integrity are enhanced with the increased number of cameras, while the cost is the texture impairments and outliers due to the multi-camera overlapping scans. However, the cost could be reduced by improving illumination conditions, which will be covered in detail in the latter discussion.    

\begin{figure}[t]
\centering
\includegraphics[width=\columnwidth]{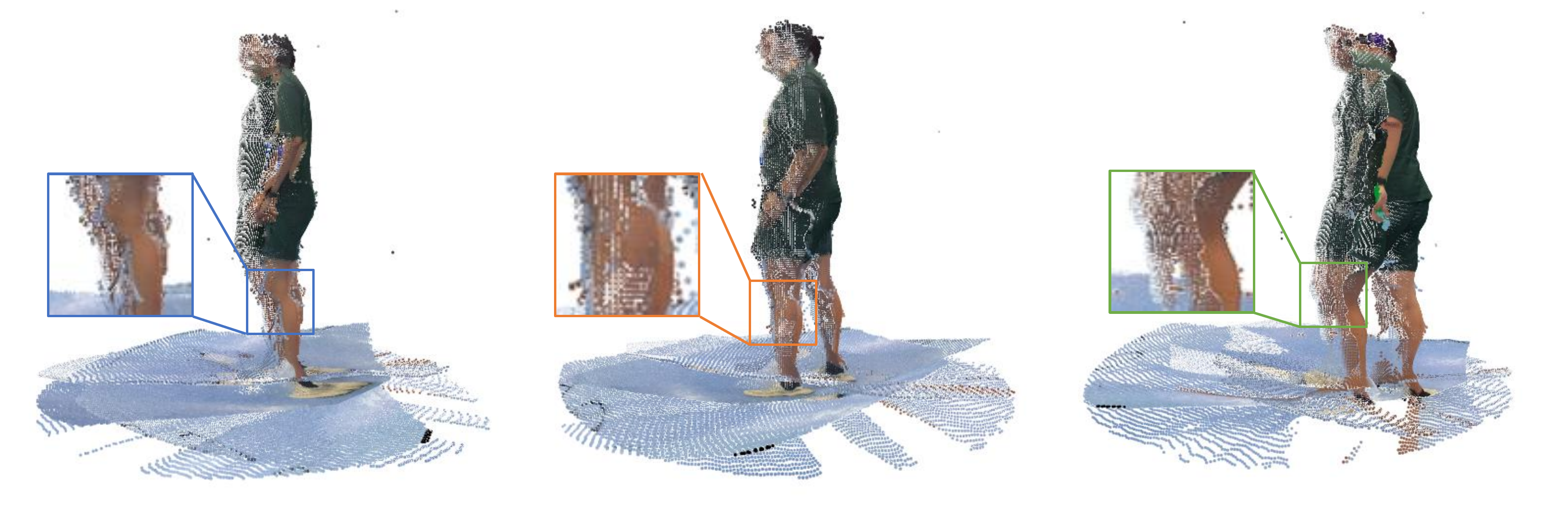}
\vspace{-20pt}
 \caption{Comparison of different amplitudes of motion
 \protect\\    Left: Static, Middle: Fidgets, Right: Walking}
 \vspace{-10pt}
\label{movement}
\end{figure}

\subsection{Movement}
Fast movement has an impact on the quality of 2D videos causing ghost images, which is also for volumetric videos. To find the impact of object movement, we conduct a contrast experiment that includes different amplitudes of motion. For the experiment setup, we make the same male actor standing in the center of the capture region and act different amplitudes of motion, with the capture duration and capture setups all the same as in the previous experiment. In experiment 1, the actor stands still in the center with no movement, in experiment 2, the actor stands still but makes fidgets occasionally, while in experiment 3, the actor walks around while waving his arms. The experiment results are depicted in Figure ~\ref{movement}, and the details in specific regions are zoomed in the box of different colors. From the results, we find that the point clouds are sparser in regions where large movements are performed, which causes the loss of detail and ghost images in specific regions.

\subsection{Scene Complexity}

\begin{figure}[t]
\centering
\includegraphics[width=\columnwidth]{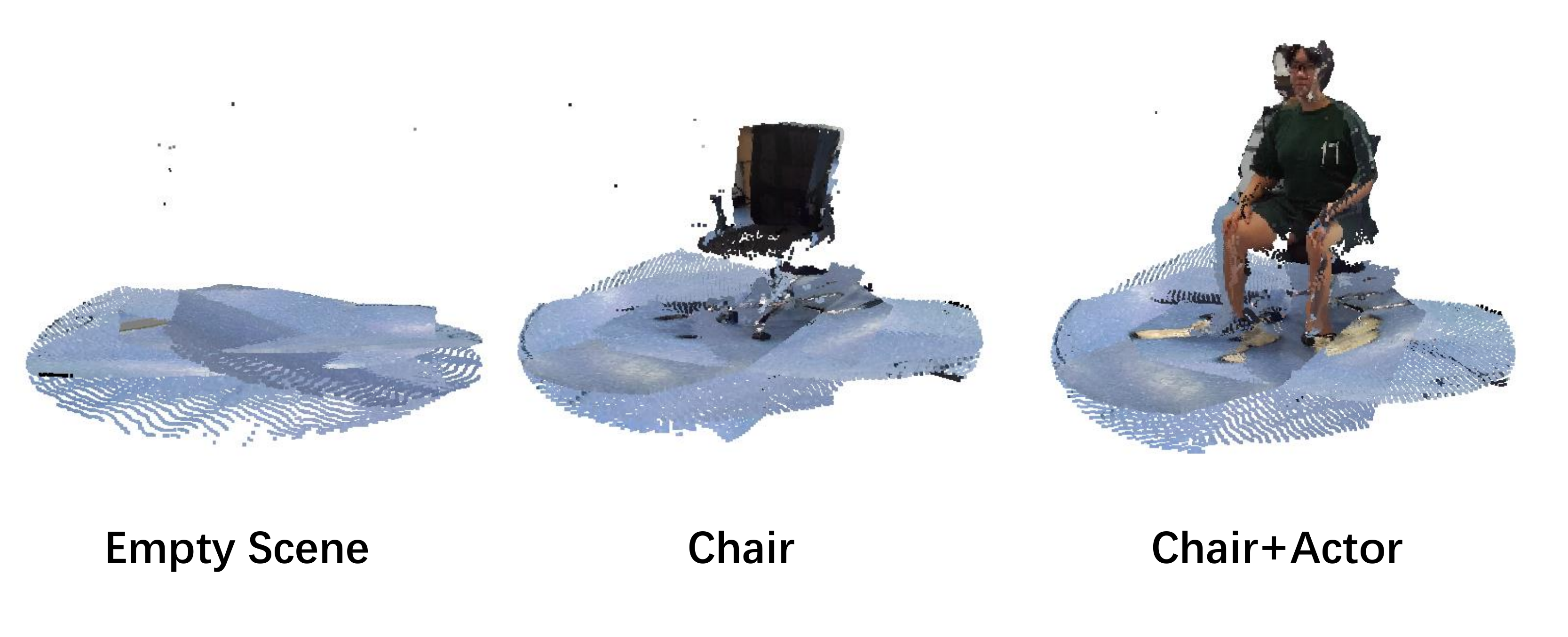}
\vspace{-30pt}
 \caption{Comparison of different scene complexity}
  \vspace{-10pt}
 \label{scene_complexity}
\end{figure}

The complexity of the scene has an impact on the size of the volumetric video data, thus we capture data in the same scene with different numbers of objects for evaluation. In such an experiment, the duration is set as 50 frames, and the capture setups are the same 6 Azure Kinects with the same illumination conditions. The scene layouts are illustrated in Figure ~\ref{scene_complexity}: For scene 1, the capture region is set to empty as a baseline. For scene 2, we add a chair in the center of the scene. And for scene 3, we add an actor sitting on a chair. Then we perform segmentation to the point clouds object by object to each of the 50 frames, as a result, we find: On average, the chair consists of 7371 points and the human consists of 16351 points. The results are listed in Table ~\ref{complexity}. According to the test result, we find that the volumetric data size increase with the complexity of the scene which contributes to the rendering costs.


\begin{table}[t]
    \centering
    \caption{Composition of different objects}
    \vspace{-10pt}
    \begin{tabular}{|l|l|l|}
    \hline
        Object & \#Point Clouds & Size \\ \hline
        Empty Scene & 58329 & 2607KB \\ \hline
        Chair & 7371 & 288KB \\ \hline
        Human & 16351 & 680KB \\ \hline

    \end{tabular}
    \label{complexity}
\end{table}

\begin{figure}[t]
\centering
\includegraphics[width=\columnwidth]{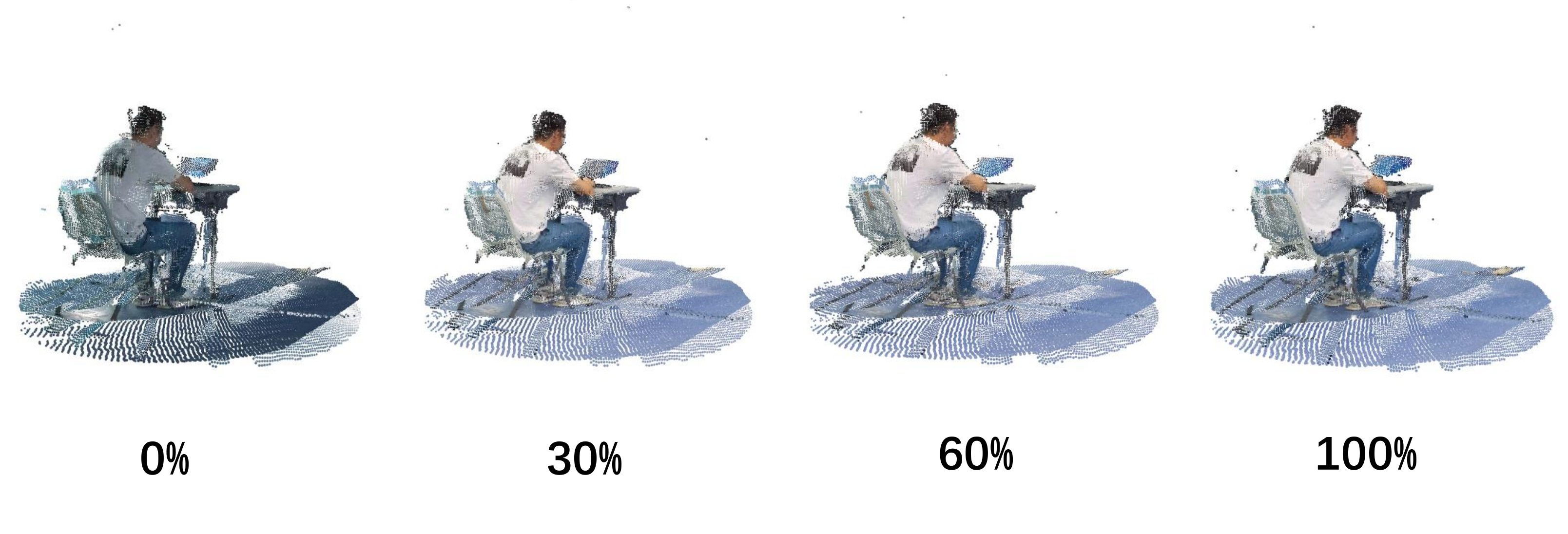}
\vspace{-30pt}
 \caption{Comparison of different illumination conditions}
  \vspace{-10pt}
 \label{illumination}
\end{figure}

\subsection{Illumination Condition}

The Microsoft Azure Kinect device integrates an RGB camera and a depth camera based on IR. The final displayable point cloud is the result of the alignment of RGB data and depth data as described in Section ~\ref{Calibration} and Section ~\ref{Alignment}. However, distortion of color information arises when the illumination condition of the capture region is limited. To evaluate the impact of illumination, we make a comparative trial under different illumination conditions. In the original capture setup, we use 2 LED lamps for illumination compensation, both are set to maximum brightness, while in this trial, we test 4 levels of illumination conditions from no external illumination to maximum brightness of the light lamps. The result is shown in Figure ~\ref{illumination}. The left figure shows the result of no external illumination, we can see a huge loss in RGB detail, in addition, the point cloud density is affected as well. The other figures show the result of 30\%, 60\%, and 100\% illumination from the light lamps. From the comparison of the results, we find loss in RGB detail and point cloud density will arise under insufficient illumination.

\section{Uses of the dataset}\label{uses}

This dataset is designed to promote the research of volumetric videos and each of the screenplays is designed with careful thought in order to provide more opportunity for algorithm optimization, parameters adjustment, and performance evaluation in aspects including capture, rendering, and transmission. 

\subsection{Virtual \& Augmented Reality}
Virtual and augmented reality are two areas where volumetric video could have a significant impact. The ability to capture and render 3D scenes with high levels of detail and realism \cite{VoluSR} could enable the creation of more immersive and engaging VR and AR experiences. For example, FSVVD could be used to create training simulations for a variety of industries, such as healthcare, military, and emergency response. These simulations could provide users with a realistic and interactive experience, allowing them to practice and learn in a safe and controllable environment.

\subsection{Social Interaction \& Communication}
In addition to training simulations, FSVVD could be used to create virtual environments for social interaction and communication. For example, the dataset could be used to create virtual worlds where users can interact with each other and with virtual objects and environments. These virtual worlds could provide a rich and engaging experience for users, allowing them to socialize, collaborate, and engage in a variety of activities.

\subsection{Healthcare}
The healthcare industry is another area where FSVVD could have a significant impact. The ability to capture and render 3D scenes with high levels of detail and realism could enable the development of machine learning models that are able to recognize and classify different medical conditions and treatments. These models could be used to assist doctors and nurses in making more accurate and timely diagnoses and treatment decisions. In addition, the dataset could be used to train models that are able to generate natural language descriptions of medical procedures and diagnoses, which could be used to improve communication and collaboration between healthcare providers and patients.

\subsection{AI Systems}
Finally, FSVVD could be used to improve the accuracy and functionality of virtual assistants and other AI-powered systems. The dataset could be used to train models that are able to understand and interact with daily life scenarios, such as recognizing and responding to different types of objects and activities in a scene. This could enable virtual assistants and other AI systems to provide more accurate and helpful assistance to users in their daily lives.

\section{Conclusions}\label{conclusions}
In conclusion, we propose FSVVD (Full Scene Volumetric Video Dataset), a publicly available volumetric video dataset that depicts human interactions with objects and full related scenes. This dataset provides more than 30 different daily scenarios and aims to provide a universal dataset for evaluation and research in the application of volumetric representation in real-life scenarios. To the best of our knowledge, FSVVD is the first volumetric dataset that provides full scenes of daily scenarios. Additionally, we perform and report a detailed analysis of the impact factors of volumetric video quality, including capture setups, object movements, scene complexity, and illumination conditions. Our future work mainly focuses on adding sequences to cover more areas and improving capture accuracy.
Overall, FSVVD contributes to energizing the research on multimedia topics of volumetric video and real-time telepresence.

\begin{acks}
The work is supported in part by the Basic Research Project No. HZQB-KCZYZ-2021067 of Hetao Shenzhen-HK S\&T Cooperation Zone, by National Natural Science Foundation of China (Grant No. 62102342), by Guangdong Basic and Applied Basic Research Foundation (Grant No. 2023A1515012668), by Shenzhen Science and Technology Program (Grant No. RCBS20221008093120047), by Shenzhen Outstanding Talents Training Fund 202002, by Guangdong Research Projects No. 2017ZT07X152 and No. 2019CX01X104, by the Guangdong Provincial Key Laboratory of Future Networks of Intelligence (Grant No. 2022B1212010001), by Young Elite Scientists Sponsorship Program by CAST (Grant No. 2022QNRC001) and by The Major Key Project of PCL Department of Broadband Communication.
\end{acks}

\bibliographystyle{ACM-Reference-Format}
\bibliography{sample-base}

\appendix

\end{document}